\documentclass{article}
\usepackage[utf8]{inputenc}
\usepackage{setspace}
\usepackage[utf8]{inputenc}
\usepackage[T1]{fontenc}
\usepackage[T2A]{fontenc}
\usepackage[russian, english]{babel}

\title{Much ado about \textit{nothing}: cosmological and anthropic limits of quantum fluctuations}
\author{Kristina Šekrst
\thanks{To appear in: \textit{Physics and Philosophy II} (ed. Luka Boršić, Dragan Poljak, Ivana Skuhala Karasman, Franjo Sokolić), Institute of Philosophy, Zagreb, 2020}}

\date{January 2020}

\begin{document}

\maketitle

\begin{quote}
  \textit{In the beginning there was nothing, which exploded.}
  \begin{flushright}
    \small{--Terry Pratchett, \textit{Lords and Ladies}}
  \end{flushright}
\end{quote}

\section{Motivation for inflationary theory}

Before the inflationary theory, the standard cosmological model had three important unsolved issues, with no answer on the horizon. The first problem is the \textit{horizon problem}: why is the universe so uniform, looking from the aspect of large-scale structures? The second question poses \textit{the flatness problem}: why is the geometry of the universe almost flat, i.e. Euclidean? Finally, the third question -- \textit{the perturbations of density problem} -- asks a seemingly philosophical question -- where did the large-scale fluctuations, which gave rise to stars and galaxies, come from?\footnote{Another important problem -- \textit{the magnetic monopole problem} -- will not be discussed here, since it carries little weight for the topic of the paper.}

``The biggest blunder of his life'', Einstein's cosmological constant, was introduced since his static space would have collapsed, so he added it to his equation in order to hold back gravity and achieve a static universe. After Hubble's discovery of the expansion of the universe, Einstein's cosmological constant was forgotten until modern times, when inflation used the option that gravity could be a repulsive force. To allow for the repulsive gravity, we require negative pressure, thus a small part of repulsive gravitational material existed in the early universe, and led to the creation of everything. Even though Einstein did not offer a physical explanation of the cosmological constant, the modern interpretation in terms of vacuum energy and pressure was introduced by George Lema\^{\i}tre \cite[p. 19]{vilenkin-many}. Therefore, the attractional gravitational force of the matter was balanced by the repulsive gravity of the vacuum. In this false vacuum, every part of the universe was of huge and constant density, and the gravitationally repulsive material, being unstable, turned into non-repulsive one. Thus the particles were produced, and afterwards the universe grew exponentially.

Alan Guth, and later Andrew Linde, proposed the theory of inflation as the exponential expansion of the early universe, which lasted somewhere from 10\textsuperscript{-35} to 10\textsuperscript{-32} seconds, when the universe expanded for 10\textsuperscript{25} times. Guth's theory aimed to resolve the conditions that led to the Big Bang. The exponential expansion of the universe -- \textit{the inflation} -- commences when a patch of the primeval chaotic \textit{quantum fluctuations} happens to expand, and the density within it drops to a point where the local energy density is dominated by the potential energy of the field, usually called \textit{inflaton}\footnote{This field was considered to be the Higgs field, however, recent discoveries are skeptical about this scenario, even though Guth himself had talked about the Higgs field
(cf. \cite[p. 175]{guth}).} \cite[p. 397]{peebles}. The energy that has been stored in the field produced high-energy particles, which collided and created other particles at high temperature, providing us with a starting point for the standard hot Big Bang cosmology \cite[p. 176]{guth}.

The inflationary theory helped cosmology to have a valid explanation for the mentioned important issues, by postulating the existence of pre-Big Bang quantum fluctuations that gave rise to an exponential expansion of the universe. To continue, regarding the horizon problem, usually different objects have similar properties if they have been in contact at some time, but according to standard cosmology, the most distant regions were not. Inflation states that these regions, which are separated today, have been in contact before the exponential growth of the universe. Even though they are at huge distances today, when they were in contact before the exponential expansion, there had been time for these regions to thermalize and reach exactly the same temperature and other physical properties \cite[p. 97]{krauss}.

The inflationary solution to the flatness problem\footnote{While \textit{having flat geometry} means that the geometry of the universe is of Euclidean type, this does not propose a \textit{flat} universe in the literal sense. It means that the shortest path between two points is a straight line, the sum of angles in a triangle is 180 degrees, and parallel lines never meet. The other two geometries have different properties: in \textit{spherical geometry}, the sum of angles in a triangle is greater than 180 degrees, and parallel lines intersect; while in \textit{hyperbolic geometry} (saddle-like), the sum of angles in a triangle is less than 180 degrees, and parallel lines are non-intersecting and move away from each other.} concerns $\Omega$, the ratio between the actual mass density of the universe and critical density.\footnote{Critical density is the density that puts the universe on the borderline -- regarding its ultimate fate -- between eternal expansion and eventual collapse.}  Recent measurement suggest a value really close to 1,\footnote{Current WMAP value is 1.0023 (+0.0056/–0.0054). Current data: NASA. 2020. “Wilkinson Microwave Anisotropy Probe”. [https://map.gsfc.nasa.gov]} which means that it started close to the value of 1, for which the standard Big Bang cosmology does not offer an explanation. Inflation reverses the problem: instead of $\Omega$ being driven away from 1, it is driven towards one exponentially, the same way as if you inflate a balloon exponentially, its surface would even out. As the inflation drives the geometry of the universe towards flatness, the value of $\Omega$ is driven to the value of 1 \cite[p. 177]{guth}.

Finally, as for the problem of perturbations of density, the inflation emphasizes the initial quantum fluctuations as seeds of today's large-scale structures: stars, galaxies, galaxy clusters, and galaxy super-clusters. The exponential expansion increased the existing quantum vacuum fluctuations, and the initial rare $\frac{1}{100000}$ uniformity had grown exponentially. The question that is immediately asked is, of course, if the inflation is here to explain the conditions that led to the Big Bang, whence do these quantum fluctuations come from?

\section{Quantum fluctuations and vacuum}

The inflationary cosmology tries to explain what happened and what ``existed'' before the Big Bang using quantum fluctuations. In quantum physics, these are temporary changes in the amount of energy in a certain point in space. The uncertainty principle allows for the temporary appearance of virtual particles out of empty space\footnote{This does not violate the conservation of energy, since the particle number operator, that counts the number of particles in systems where the total amount of particles may not
be preserved, lacks the commutative property with a field’s Hamiltonian operator -- the lowest energy state or the vacuum state.} \cite{tamm}. Since there is not a precise determination of position and momentum, or energy and time, a particle pair may pop out of the vacuum in a short interval. Therefore, it is possible for a particle and its antiparticle -- say, an electron and a positron -- to materialize from the vacuum, exist briefly, and disappear. In this case, the separation between the electron and the positron is typically no larger than 10\textsuperscript{-10} cm, and the fluctuation typically lasts around 10\textsuperscript{-21} seconds, so even though they cannot be observed directly, it can be done indirectly. For example, atomic physicists can usually calculate the magnetic strength of an electron up to ten decimal places, and if the influence of the materialization of the electron-positron pairs is left out, the answer agrees with the experiment only for the first five decimal places, while the next five turn out wrong \cite[p. 272]{guth}.

In the core of inflationary theory lie the mentioned quantum fluctuations. Since the Big Bang marks the creation of everything: space, time, and matter, the question that is immediately posed is: \textit{what exactly did fluctuate}? If inflationary theory resolves cosmological issues by stating that there were quantum fluctuations and virtual particles, this would imply that there was a certain something, not at all nothing before the Big Bang.\footnote{This is common criticism against Lawrence Krauss \cite{krauss}.}

\section{Universe as an inflated quantum fluctuation}

According to the first law of thermodynamics, nothing in the universe cannot pop into existence out of nothing. The question can the universe itself do that was posed by Edward Tryon in 1970,\footnote{Tryon was at Dennis Sciama’s lecture at the Columbia University when he proposed this, but everyone thought it was a joke \cite[p. 183]{vilenkin-many}.} considering the universe as a quantum fluctuation \cite{tryon}. In principle any object might materialize briefly in the vacuum,\footnote{Guth \cite[p. 272]{guth} jokingly mentions a refrigerator or a pocket calculator. \textit{Caveat}:
the conservation laws cannot be violated, and if the object is positively charged, then it can materialize only if the equal amount of negative charge is produced.} but the probability, of course, significantly decreases if the object is more massive and more complex. However, it is still a probabilistic issue, end even though it is highly improbable we should get anything other than a virtual particle pair, it could be possible that whole universes might materialize from the vacuum, including ours.

The problem is: the lifetime of the fluctuation depends on its mass, and the lifespan is shortened as the mass increases, so for a quantum fluctuation to survive for more than ten billion years, the mass of it should be less than 10\textsuperscript{-65} grams \cite[p. 272]{guth}. However, in a closed universe, the negative gravitational energy cancels the energy of matter, and the total mass is equal to zero, so the lifetime of a such quantum fluctuation can be infinite. This scenario presupposes quantum fields in some kind of space (and the standard model attributes the creation of space and time to the Big Bang itself), so the question is how to reformulate this idea without spacetime.

Vilenkin \cite{vilenkin} took an empty geometry, not a closed one, and the concept of quantum tunneling\footnote{\textit{Quantum tunneling} is a quantum mechanical phenomenon in which a particle tunnels through a barrier that it should not be able to move through in classical terms, i.e. it does not have enough energy to move over it. However, in quantum mechanics, particles behave as waves as well, so if the barrier is thin enough, the probability function describing
a wave may extend to the next region over the barrier.} to conceive the idea that the universe started from an \textit{empty geometry}, containing no points, and then tunneled into a non-empty state. That way, the problem of where the fluctuations happened conceptually disappears, and we can use the theory of inflation to increase the fluctuations to today's size.

Thus, we use inflationary theory to explain how the universe appeared out of nothing, starting from quantum fluctuations that tunneled from an empty state to a non-empty one, and this subatomic universe thus created was enlarged accordingly by the exponential expansion. The problem is that we still start with something, since the particles that can jump or tunnel through an energy barrier must still initially exist -- even for a brief flash of time -- in order to do so. From a philosophical point of view, declaring the initial condition to be an empty geometry still presupposes \textit{something} having the no-points properties. So, if the universe owes its origin to the quantum theory combined with the cosmological model of inflation, it seems that the quantum laws must have existed before the creation of the universe, which again motivates us to ask the question what led to these quantum fluctuations.

\section{Anthropic limitations}

One of philosophically interesting principles that can be used to clarify this issue is the anthropic principle (proposed by Barrow and Tipler \cite{barrow-tipler}): the universe is how it is since it allows for us to observe it, i.e. only the universe that is capable of supporting life will have beings capable of reflecting on its fine tuning. Combined with the possibility of creation of more than one universe -- no matter how statistically improbable -- there would be some universes that could tunnel and inflate to a larger size.\footnote{The most likely to be created would be Planck-sized universes which would instantly recollapse, and the observers could not be involved in these.}

The issues with all theories that presuppose various fluctuations and energy levels, combined with inflation in order to achieve the current state, is that they have to presuppose some properties and entities such as \textit{particles}, \textit{energy}, \textit{tunneling} etc. in order to explain the creation of the universe. If one takes into account the anthropic principle, one could describe our need to explain the notions of \textit{nothingness}, \textit{vacuum fluctuations}, \textit{quantum chaos} as a way of observing our current universe from the only aspect we know: the aspect of \textit{something}. One possibility is that physics could never thoroughly explain the nature of quantum fluctuations, even if we did know that our universe happened to be one of the possible inflated fluctuations, because our observing apparatus is necessarily linked to the universe made out of matter, presupposing something to exist, without a possible existence of an empty region devoid of matter, energy, and quantum fluctuations.

\section{Computational limitations}

One of the biggest issues in cosmology is the problem of experiments that cannot be repeated -- such as the creation of the universe -- and the problem of verifying events that have happened out of our reach. Even if we did have a complete theory of everything, it may still not be able to give us the correct answer, but it could give us the underlying principle. Recent tries of simulating virtual particles have established that we could simulate high-energy physics, showing how they could behave at energy levels too high to easily generate them in reality. Martinez et al. \cite{martinez} used a quantum computer built using four electromagnetically trapped calcium ions, whose spins were used as qubits, and were controlled by laser pulses to perform logic operations. A pair of qubits\footnote{In quantum computing, a qubit is a unit of quantum information, which differs from the classical bit that has to be in one state or the other since quantum mechanics allows it to be in a superposition of both states simultaneously.} represented a pair of virtual particles, and the resulting quantum fluctuations in energy allowed them to read off whether particles and their anti-particle pairs were created in a simulation.

The question of could we simulate the quantum fluctuations before the Big Bang seems to be connected with the rising complexity, since interactions between boundary particles cause an explosive growth in the complexity of their collective quantum state. Computational complexity observes how much resources we need in order to perform a certain operation, and for some problems, we would need resources larger than the entire universe has to offer \cite[p. 264]{aaronson}. In computational complexity, there are \textbf{NP} problems, for whose resolution we would need more resources that the universe can handle, but are easily verifiable.\footnote{For example, it is easy to use a password for a certain login, but it is not as easy to crack one. The most famous question today is the \textbf{P} versus \textbf{NP} problem that asks can we reduce the quickly verifiable problems (\textbf{NP}) to those that are quickly solved (\textbf{P}) by a computer. The problems is still unresolved, but the majority of researchers believe that \textbf{NP} problems cannot be reduced to \textbf{P} problems.} If there is a link between the growth of the computational complexity and the expansion of the universe, the computational complexity may have a key role in the complete theory of quantum gravity and the beginning of the universe.

If the simulation of the pre-Big Bang conditions would be shown to be verifiable, but unsolvable, that gives us hope for the future simulations of these initial conditions, since we could test the complete theory and see whether the answer checks out. That way, the problem of verification and refutation in this aspect of cosmology should disappear.

Unfortunately, if we do not develop a theory of everything, the answer should forever stay out of our reach, where one could pose a link between the anthropic principle and our ability to produce such a computational power. If we are not able to simulate the initial conditions, since the resources we need should ask for more than this universe can provide us with, and there is no way to reach the other possible universes\footnote{Some inflationary theories presuppose various inflated bubbles that lead to the creation of various universes. However, since these expand at the speed of light, and particles of our type cannot exist in these universes, such regions will always stay out of our observing capabilities.} -- maybe the computational limit itself is a part of our anthropic reasoning condition, and the universe itself puts a constraint to explain its beginning. Thus, we can use quantum mechanics and inflation to hypothesize about their development, including the creation of the universe as a result. However, the creation itself could have closed the door to the understanding of the essence of the \textit{nothingness} in the beginning.

\bibliography{sekrst}
\bibliographystyle{acm}

\end{document}